\documentclass[prb,12pt]{revtex4-1}

\usepackage[english]{babel}
\usepackage[intlimits]{amsmath}
\usepackage[dvips]{graphicx}
\usepackage{amsfonts}
\usepackage{amssymb}
\usepackage{graphicx}
\usepackage{bm}

\begin{document}

\begin{titlepage}

\title{All-time dynamics of continuous-time random walks on complex networks}
\author {Hamid Teimouri, Anatoly B. Kolomeisky}
\affiliation{Department of Chemistry, Rice University, Houston, TX 77005-1892}

\begin{abstract}

The concept of continuous-time random walks (CTRW) is a generalization of ordinary random walk models, and it is a powerful tool for investigating a broad spectrum of phenomena in natural, engineering, social and economic sciences. Recently, several theoretical approaches have been developed that allowed to analyze explicitly dynamics of CTRW at all times, which is critically important for understanding mechanisms of underlying phenomena. However, theoretical analysis has been done mostly for systems with a simple  geometry. Here we extend the original method based on generalized master equations to analyze all-time dynamics of CTRW models on complex networks. Specific calculations are performed for models on lattices with branches and for models on coupled parallel-chain lattices. Exact expressions for velocities and dispersions are obtained. Generalized fluctuations theorems for CTRW models on complex networks are discussed.

\end{abstract}

\maketitle

\end{titlepage}
\section{Introduction}

Continuous-time random walks (CTRW) are stochastic lattice models that have been introduced by Montroll and Weiss in 1965 as a generalization of ordinary random walk processes.\cite{montroll65,montroll73,landman77} In these models the motion of particles between different states is generally controlled by random waiting-time distribution functions. Simple random walks are recovered when the waiting-time distributions are exponential functions.\cite{montroll65,montroll73,landman77} It turned out that CTRW is a powerful and efficient method for studying a wide distribution of complex dynamic processes in natural sciences, engineering, social sciences and in economics.\cite{weiss_book,hughes_book,rudnick_book,metzler00,zaslavsky02,econoCTRW,saxton12,denz12,kang11,sippel11} It is natural to utilize CTRW for understanding transport phenomena that cannot be described within classical diffusion framework.\cite{montroll65,montroll73,metzler00} Recent experimental advances in single-molecule techniques that allowed to visualize various chemical and biological processes with high temporal and spatial resolution stimulated the application of CTRW for investigating biological and cellular transport phenomena with anomalous diffusion.\cite{min05,kolomeisky00,AR,dagdug09,leijnse12}

Theoretical studies of complex dynamic phenomena that utilize CTRW models mostly concentrate on large-time dynamics.\cite{montroll65,weiss_book} While this might be a reasonable approach for systems where stationary states are well established and can be quickly reached, in many cases to fully understand mechanisms of complex processes one needs to have a description of dynamic behavior at all times. In addition, the situation is more complex for system that never reach the steady-state limit. The original work of Montroll and Weiss\cite{montroll65} suggested that all-time dynamics of CTRW can be obtained by utilizing Laplace-Fourier transformations. Although this analysis is formally correct, it is practically impossible to apply for real dynamic processes. Recently, a new method of calculating dynamic properties of particles in the CTRW model at all times has been introduced.\cite{BW08} It is based on analyzing a propagator and surviving probabilities to calculate analytically exactly Laplace transforms of all dynamic properties, which effectively  provides  all-time description of the underlying processes. It was later extended to a CTRW model with branched states as a way of investigating biased diffusion in tubes with periodic dead ends.\cite{berezhkovskii11}  However, it can be shown that this method could be utilized only for homogeneous CTRW with the same set of waiting time distributions at each site, limiting its application for more complex dynamic phenomena. An alternative theoretical method of calculating all-time dynamic properties in CTRW models have been presented by one of the authors.\cite{kolomeisky09} It utilizes a generalized master-equations approach\cite{kenkre73,kolomeisky00} that allows fast and efficient computation of all Laplace transforms of probability functions and all dynamic properties of the CTRW models. The advantage of the method is the ability to analyze some inhomogeneous CTRW models as well as more complex systems, as was shown explicitly for simple periodic CTRW models and for processes with irreversible detachments.\cite{kolomeisky09}

Most of chemical and biological processes can be viewed as complex networks of states that are connected by dynamic transitions. The application of CTRW method for understanding mechanisms of these systems require a theoretical analysis that is valid at all time scales. In this paper, we extend the generalized master-equations method to analyze all-time dynamics of CTRW models on complex networks. It is interesting to note that closely related method has been proposed recently for temporal fully-connected networks.\cite{hoffmann12} In this work, stationary networks will be investigated. Specifically, dynamic properties for the CTRW models with branched states and the coupled parallel-chain CTRW models are obtained and analyzed. In addition, generalized fluctuations theorems for these systems are discussed.

\section{ CTRW on Lattices with Branches}

There are many different type of geometries in complex networks that might characterize various dynamic processes. In this paper, we specifically consider 2 types of geometries: a model with branched states and a parallel-chain lattice model as shown in Fig. 1; but the analysis can also be straightforwardly  extended to other network topologies.

First, we  consider the CTRW model on the lattice with branched states, as illustrated in Fig. 1a. The dynamics of the random walker in this system is governed by a set of waiting-time distribution functions. The particle at the site $i$ on the main lattice can jump one step  forward with the probability $\psi_{i}^{+}(t)dt$, while the step backward can take place with the probability $\psi_{i}^{-}(t)dt$. The particle could also move into the branched state with the probability  $\psi_{i}^{\beta}(t)dt$, and the motion back to the main lattice is controlled by the probability $\psi_{i}^{\gamma}(t)dt$ (see Fig. 1a). It is assumed that the system is homogeneous, i.e., waiting-time distribution function are independent of the state $i$: $\psi_{i}^{\pm}(t)= \psi^{\pm}(t)$;  $\psi_{i}^{\beta}(t)= \psi^{\beta}(t)$ and $\psi_{i}^{\gamma}(t)= \psi^{\gamma}(t)$. Since detachments into the branched state are reversible the probability to find the particle in the system is conserved.

Our method of calculating all-time dynamic properties in CTRW models is based on the important result obtained by Landman, Montroll and Shlesinger in 1977.\cite{landman77} They showed that the dynamics of the particle in any CTRW model can be fully described by a generalized master equation. For the system with branched states we define $P_{n,j}(t)$ as the probability of finding the random walker at the site $n$ at time $t$. Here, the label $j=0$ corresponds to the site on the main lattice, while $j=1$ represents the branched state. It is assumed that at $t=0$ the particle starts at the origin ($n=0$) on the main lattice. Then the corresponding generalized master equations  can be written as
\begin{eqnarray}\label{master_beq1}
 \frac{d P_{n,0}(t)}{dt}=\int_{0}^{t} [\varphi^{+}(\tau)P_{n-1,0}(t-\tau)+\varphi^{-}(\tau)P_{n+1,0}(t-\tau) + \varphi^{\gamma}(\tau)P_{n,1}(t-\tau)] d\tau \nonumber \\ 
 - \int_{0}^{t} [\varphi^{+}(\tau) + \varphi^{-}(\tau)+ \varphi^{\beta}(\tau)] P_{n,0}(t-\tau) d\tau;
\end{eqnarray}
\begin{equation}\label{master_beq2}
\frac{d P_{n,1}(t)}{dt}=\int_{0}^{t} \left\{ \varphi^{\beta}(\tau) P_{n,0}(t-\tau) - \varphi^{\gamma}(\tau)  P_{n,1}(t-\tau)  \right\} d\tau.
\end{equation}
Here we used waiting-time rate distributions $\varphi^{\pm}(t)$, $\varphi^{\beta}(t)$ and $\varphi^{\gamma}(t)$, and it can be shown that they are related to waiting-time distribution functions via Laplace transforms,\cite{kolomeisky00,kolomeisky09}
\begin{equation}\label{transformb}
\widetilde{\varphi}^{\pm}(s)=\frac{s \widetilde{\psi}^{\pm}(s)}{1-\widetilde{\psi}(s)}, \quad 
\widetilde{\varphi}^{\beta}(s)=\frac{s \widetilde{\psi}^{\beta}(s)}{1-\widetilde{\psi}(s)}, \quad
\widetilde{\varphi}^{\gamma}(s)=\frac{s \widetilde{\psi}^{\gamma}(s)}{1-\widetilde{\psi^{\gamma}}(s)},
\end{equation}
with $\widetilde{\psi}(s)=\widetilde{\psi}^{+}(s)+\widetilde{\psi}^{-}(s)+\widetilde{\psi}^{\beta}(s)$.

After performing Laplace transformations the generalized master equations are modified into 
\begin{equation} \label{recurb1}
\left[s+\widetilde{\varphi}^{+}(s)+\widetilde{\varphi}^{-}(s)+\widetilde{\varphi}^{\beta}(s)\right] \widetilde{P}_{n,0}(s)=\delta_{n,0}+ \widetilde{\varphi}^{+}(s)\widetilde{P}_{n-1,0}(s) + \widetilde{\varphi}^{-}(s)\widetilde{P}_{n+1,0}(s) + \widetilde{\varphi}^{\gamma}(s)\widetilde{P}_{n,1}(s);
\end{equation}
\begin{equation} \label{recurb2}
\left[s+\widetilde{\varphi}^{\gamma}(s)\right] \widetilde{P}_{n,1}(s)= \widetilde{\varphi}^{\beta}(s)\widetilde{P}_{n,0}(s). 
\end{equation}
It is convenient to introduce new auxiliary variables: $a \equiv s+\widetilde{\varphi}^{+}(s)+\widetilde{\varphi}^{-}(s)+\frac{s  \widetilde{\varphi}^{\beta}(s) }{s+ \widetilde{\varphi}^{\gamma}(s)}$, $b \equiv \widetilde{\varphi}^{+}(s)$, $c \equiv \widetilde{\varphi}^{-}(s)$ and $d \equiv \frac{\widetilde{\varphi}^{\beta}(s)}{s+\widetilde{\varphi}^{\gamma}(s)}$.  Combining Eqs.(\ref{recurb1}) and (\ref{recurb2}) and utilizing these variables leads to a simpler expression,
\begin{equation}\label{eqB}
a \widetilde{P}_{n,0}(s)=\delta_{n,0}+ b \widetilde{P}_{n-1,0}(s) + c \widetilde{P}_{n+1,0}(s).
 \end{equation}
This recursion relation can be  solved exactly,\cite{kolomeisky09} and the solution is given by:
\begin{equation}
\widetilde{P}_{n,0}(s) = \left(\frac{b}{c}\right)^{n/2} \left( \frac{ a- \sqrt{a^{2}-4 bc}}{2 \sqrt{bc}}\right)^{|n|} \frac{1}{\sqrt{a^{2}-4 bc}},
\end{equation}
or more explicitly,
\begin{eqnarray}\label{eq_Pn_phi}
& \widetilde{P}_{n,0}(s) = \left(\frac{\widetilde{\varphi}^{+}(s)}{\widetilde{\varphi}^{-}(s)} \right)^{n/2}  \left( \frac{ s+\widetilde{\varphi}^{+}(s)+\widetilde{\varphi}^{-}(s)
+\frac{s\widetilde{\varphi}^{\beta}(s)}{s+\widetilde{\varphi}^{\gamma}(s)} - \sqrt{[s+\widetilde{\varphi}^{+}(s)+\widetilde{\varphi}^{-}(s)+\frac{s\widetilde{\varphi}^{\beta}(s)}{s+\widetilde{\varphi}^{\gamma}(s)}]^{2}-4 \widetilde{\varphi}^{+}(s)\widetilde{\varphi}^{-}(s)}}{2 \sqrt{\widetilde{\varphi}^{+}(s)\widetilde{\varphi}^{-}(s)}} \right)^{|n|} \times \nonumber \\
& \frac{1}{\sqrt{[s+\widetilde{\varphi}^{+}(s)+\widetilde{\varphi}^{-}(s)+\frac{s\widetilde{\varphi}^{\beta}(s)}{s+\widetilde{\varphi}^{\gamma}(s)}]^{2}-4 \widetilde{\varphi}^{+}(s) \widetilde{\varphi}^{-}(s)}}.
\end{eqnarray}
It can also be written in terms of the original waiting-time distribution functions with the help of Eqs. (\ref{transformb}),
\begin{eqnarray}\label{eq_Pn}
& &\widetilde{P}_{n,0}(s) = \left(\frac{\widetilde{\psi}^{+}(s)}{\widetilde{\psi}^{-}(s)} \right)^{n/2} \left( \frac{ 2\sqrt{\widetilde{\psi}^{+}(s)\widetilde{\psi}^{-}(s)}}{1-\widetilde{\psi}^{\beta}(s)\widetilde{\psi}^{\gamma}(s)+ \sqrt{(1-\widetilde{\psi}^{\beta}(s)\widetilde{\psi}^{\gamma}(s))^{2}-4\widetilde{\psi}^{+}(s)\widetilde{\psi}^{-}(s)}} \right)^{|n|} \times \nonumber\\
& & \frac{(1-\widetilde{\psi}(s))/s}{\sqrt{(1-\widetilde{\psi}^{\beta}(s)\widetilde{\psi}^{\gamma}(s))^{2}-4\widetilde{\psi}^{+}(s)\widetilde{\psi}^{-}(s)}}.
\end{eqnarray}
Similar formulas for $\widetilde{P}_{n,1}(s)$ are easily obtained directly from Eq. (\ref{recurb2}). These results are exactly the same as have been obtained in Ref. 20 by a different method.

Exact analytical expressions for Laplace transforms of probability distribution functions  [Eqs. (\ref{eq_Pn})] provide a convenient method of analyzing dynamics of CTRW model on the lattice with branched states at all times. To illustrate this, let us calculate first two moments of the motion for the system. We define $\langle n(t) \rangle$ as the average position of the random walker at time $t$, and the corresponding Laplace transform is given by
\begin{equation}\label{eqbns1}
\langle \widetilde{n}(s) \rangle= \sum_{n=-\infty}^{\infty} n( \widetilde{P}_{n,0}(s)+ \widetilde{P}_{n,1}(s) ) = (1+d) \sum_{n=-\infty}^{\infty} n \widetilde{P}_{n,0}(s),
\end{equation}
which yields\cite{kolomeisky09}
\begin{eqnarray}\label{eqbns}
\langle \widetilde{n}(s) \rangle=(1+d) \frac{(b-c)}{(a-b-c)^{2}}= \frac{1} {1+ \frac{ \widetilde{\varphi}^{\beta}(s) }{s+ \widetilde{\varphi}^{\gamma}(s)}}[ \frac{\widetilde{\varphi}^{+} (s) - \widetilde{\varphi}^{-}(s) } {s^{2}}].
\end{eqnarray}
Similar analysis of the Laplace transform for the second moment produces,
\begin{eqnarray}\label{eqbn2s}
& &\langle \widetilde{n}^{2}(s) \rangle= \sum_{n=-\infty}^{\infty} n^{2}( \widetilde{P}_{n,0}(s) + \widetilde{P}_{n,1}(s) ) = (1+d)\sum_{n=-\infty}^{\infty} n^{2} \widetilde{P}_{n,0}(s)  \nonumber \\
&&= (1+d) \frac{(b+c)(a+b+c)-8bc}{(a-b-c)^{3}} \nonumber \\
& &= \frac{[\widetilde{\varphi}^{+} (s) + \widetilde{\varphi}^{-}(s)][s+2\widetilde{\varphi}^{+} (s) + 2\widetilde{\varphi}^{-}(s)+ \frac{s \widetilde{\varphi}^{\beta}(s)}{s+ \widetilde{\varphi}^{\gamma}(s)}]-8\widetilde{\varphi}^{+} (s) \widetilde{\varphi}^{-}(s) } {s^{3} (1+ \frac{ \widetilde{\varphi}^{\beta}(s)}{s+ \widetilde{\varphi}^{\gamma}(s)})^2}.
\end{eqnarray}

The steady-state dynamic behavior of the first and second moments ($t \rightarrow \infty$) can be found by considering the limit of $s \rightarrow 0$. Expanding Laplace transforms of waiting-time rate distributions for small $s$, one could write\cite{kolomeisky00}
\begin{eqnarray}
\widetilde{\varphi}^{+}(s) \simeq u + g^{+}s + \cdots,  \quad  \widetilde{\varphi}^{-}(s) \simeq w + g^{-}s + \cdots,  \nonumber \\
\widetilde{\varphi}^{\beta}(s) \simeq \beta + g^{\beta}s + \cdots, \quad  \widetilde{\varphi}^{\gamma}(s) \simeq \gamma + g^{\gamma}s + \cdots, 
\end{eqnarray}  
where $u=\widetilde{\varphi}^{+}(s=0)$, $w=\widetilde{\varphi}^{-}(s=0)$, $\beta=\widetilde{\varphi}^{\beta}(s=0)$ and $\gamma=\widetilde{\varphi}^{\gamma}(s=0)$ are effective transition rates;\cite{kolomeisky00} while $g^{\pm}=\frac{d\widetilde{\varphi}^{\pm}}{ds}|_{s=0}$, $g^{\beta}=\frac{d\widetilde{\varphi}^{\beta}}{ds}|_{s=0}$ and $g^{\gamma}=\frac{d\widetilde{\varphi}^{\gamma}}{ds}|_{s=0}$. To proceed further, we substitute these expansions into Eqs. (\ref{eqbns}) and (\ref{eqbn2s}), leading to
\begin{equation}
 \langle n(t) \rangle \simeq \frac{(u-w)}{1+\frac{\beta}{\gamma}} t + \frac{(g^{+}-g^{-})}{1+\frac{\beta}{\gamma}}+\frac{(u-w)}{(1+\frac{\beta}{\gamma})^2}\left(\frac{1+g^{\gamma}}{\gamma}-\frac{g^{\beta}}{\beta}\right)\frac{\beta}{\gamma},
\end{equation}
\begin{equation}
 \langle n^2(t) \rangle \simeq \frac{(u-w)^2}{(1+\frac{\beta}{\gamma})^2} t^2 +\left[\frac{4(u-w)(g^{+}-g^{-})}{(1+\frac{\beta}{\gamma})^2}+ \frac{(u+w)}{(1+\frac{\beta}{\gamma})}+\frac{4(u-w)^2}{(1+\frac{\beta}{\gamma})^3}\left(\frac{1+g^{\gamma}}{\gamma}-\frac{g^{\beta}}{\beta}\right)\frac{\beta}{\gamma}\right]t.
\end{equation}

The large-time behavior of the moments  allows us to compute  important dynamic properties such as the effective drift velocity $V$ and the effective diffusion constant $D$. The velocity is found from
\begin{equation}
V=\lim_{t \rightarrow \infty} \frac{ d \langle n(t) \rangle }{dt}= \frac{(u-w)}{1+\frac{\beta}{\gamma}}.
\end{equation}
The corresponding expression for the diffusion constant is given by
\begin{equation}
D=\frac{1}{2} \lim_{t \rightarrow \infty} \frac{ d \langle n^{2}(t) \rangle - \langle n(t) \rangle^{2}}{dt}= \frac{(u+w)}{(1+\frac{\beta}{\gamma})}+ \frac{(u-w)(g^{+}-g^{-})}{(1+\frac{\beta}{\gamma})^2}+\frac{(u-w)^2}{(1+\frac{\beta}{\gamma})^3}\left(\frac{1+g^{\gamma}}{\gamma}-\frac{g^{\beta}}{\beta}\right)\frac{\beta}{\gamma}.
\end{equation}
It is important to note that these equations reproduce expressions for stationary-state dynamic properties of the CTRW model with  branched states, as was obtained earlier by different methods.\cite{kolomeisky00}

\section{CTRW on Coupled Parallel-Chain Lattices}

In this section, a CTRW model on more complex network that consists of coupled parallel chains (see Fig. 1b) is investigated. The random walker can be found in one of two lattices. On the lattice 1, the waiting time distributions to move forward or backward along the same chain are given by probabilities $\psi_{1}^{+}(t)dt$ and  $\psi_{1}^{-}(t)dt$, respectively. Similarly, $\psi_{2}^{\pm}(t)dt$ are the probabilities to hop along the lattice 2: see Fig. 1b. The particle from the lattice 1 can jump to the same site on another channel with the probability $\psi^{\gamma}(t)dt$, while the reversed motion has the probability $\psi^{\delta}(t)dt$. Since only the homogeneous CTRW model is analyzed here these probabilities do not depend on the site position. Again, we  introduce a function $P_{n,i}(t)$ as the probability to find the particle on the site $n$ on the channel $i$ ($i=1,2$) at time $t$. For initial conditions we assume that the particle at $t=0$ starts in the first channel at the site $n=0$. Then the temporal evolution of the system is governed by a set of generalized master equations:  
\begin{eqnarray}\label{parallel_eq1}
 \frac{d P_{n,1}(t)}{dt}=\int_{0}^{t} [{ \varphi_{1}^{+}(\tau) P_{n-1,1}(t-\tau)+ \varphi_{1}^{-}(\tau) P_{n+1,1}(t-\tau)+ \varphi^{\delta}(\tau) P_{n,2}(t-\tau)}]d\tau, \nonumber \\
 -\int_{0}^{t}[\varphi_{1}^{+}(\tau)+\varphi_{1}^{-}(\tau) + \varphi^{\gamma}(\tau)]P_{n,1}(t-\tau) d\tau
\end{eqnarray}
\begin{eqnarray}\label{parallel_eq2}
 \frac{d P_{n,2}(t)}{dt}=\int_{0}^{t} [{ \varphi_{2}^{+}(\tau) P_{n-1,2}(t-\tau) + \varphi_{2}^{-}(\tau) P_{n+1,2}(t-\tau)  + \varphi^{\gamma}(\tau) P_{n,1}(t-\tau)}] d\tau  \nonumber \\
 - \int_{0}^{t}{[\varphi_{2}^{+}(\tau) + \varphi_{2}^{-}(\tau) + \varphi^{\delta}(\tau)]P_{n,2}(t-\tau)} d\tau.
\end{eqnarray}
In these equations rate-distribution functions are related to waiting-time distributions as follows:
\begin{eqnarray}\label{transform}
 \widetilde{\varphi_{1}}^{\pm}(s)=\frac{s \widetilde{\psi_{1}}^{\pm}(s)}{1-\widetilde{\psi_{1}}(s)}; \quad  \widetilde{\varphi}^{\gamma}(s)=\frac{s \widetilde{\psi}^{\gamma}(s)}{1-\widetilde{\psi_{1}}(s)}; \nonumber \\
 \widetilde{\varphi_{2}}^{\pm}(s)=\frac{s \widetilde{\psi_{2}}^{\pm}(s)}{1-\widetilde{\psi_{2}}(s)}; \quad \widetilde{\varphi}^{\delta}(s)=\frac{s \widetilde{\psi}^{\delta}(s)}{1-\widetilde{\psi_{2}}(s)};
\end{eqnarray}
with $\widetilde{\psi_{1}}(s)=\widetilde{\psi_{1}}^{+}(s)+\widetilde{\psi_{1}}^{-}(s)+\widetilde{\psi}^{\gamma}(s)$ and  $\widetilde{\psi_2}(s)=\widetilde{\psi_{2}}^{+}(s)+\widetilde{\psi_{2}}^{-}(s)+\widetilde{\psi}^{\delta}(s)$.

Applying the Laplace transformation to generalized master equations, we obtain,
 \begin{equation} \label{recurp1}
\left[s+\widetilde{\varphi_{1}}^{+}(s)+\widetilde{\varphi_{1}}^{-}(s)+\widetilde{\varphi}^{\gamma}(s)\right] \widetilde{P}_{n,1}(s)=\delta_{n,0}+ \widetilde{\varphi_{1}}^{+}(s)\widetilde{P}_{n-1,1}(s) + \widetilde{\varphi_{1}}^{-}(s)\widetilde{P}_{n+1,1}(s) + \widetilde{\varphi}^{\delta}(s)\widetilde{P}_{n,2}(s). 
\end{equation}
 \begin{equation} \label{recurp2}
\left[s+\widetilde{\varphi_{2}}^{+}(s)+\widetilde{\varphi_{2}}^{-}(s)+\widetilde{\varphi}^{\delta}(s)\right] \widetilde{P}_{n,2}(s)= \widetilde{\varphi_{2}}^{+}(s)\widetilde{P}_{n-1,2}(s) + \widetilde{\varphi_{2}}^{-}(s)\widetilde{P}_{n+1,2}(s) + \widetilde{\varphi}^{\gamma}(s)\widetilde{P}_{n,1}(s). \end{equation}
The conservation of probability in this system, $\sum_{n=-\infty}^{\infty} \left[ P_{n,1}(t)+P_{n,2}(t) \right] =1$, leads to the following expression in the Laplace space,
\begin{equation}\label{normcondb}
 \sum_{n=-\infty}^{\infty} \left[ \widetilde{P}_{n,1}(s)+ \widetilde{P}_{n,2}(s) \right] = \frac{1}{s}
\end{equation}
 For convenience, we define new parameters: $d_{1}\equiv \widetilde{\varphi^{\gamma}}(s)$, $d_{2}\equiv \widetilde{\varphi^{\delta}}(s)$, $ b_{1}\equiv \widetilde{\varphi_{1}}^{+}(s)$, $ c_{1}\equiv \widetilde{\varphi_{1}}^{-}(s)$, $ b_{2}\equiv \widetilde{\varphi_{2}}^{+}(s)$ and $ c_{2}\equiv \widetilde{\varphi_{2}}^{-}(s)$. Then summing over  Eqs. (\ref{recurp1}) and (\ref{recurp2}) produces
\begin{eqnarray}\label{normp1}
 \sum_{n=-\infty}^{\infty} \widetilde{P}_{n,1}(s)= \frac{d_{2}}{s+d_{1}}\sum_{n=-\infty}^{\infty} \widetilde{P}_{n,2}(s)),
\end{eqnarray}
\begin{eqnarray}\label{normp2}
 \sum_{n=-\infty}^{\infty} \widetilde{P}_{n,1}(s)= \frac{s+d_{2}}{d_{1}}\sum_{n=-\infty}^{\infty} \widetilde{P}_{n,2}(s)), 
\end{eqnarray}
which suggests that the following relation holds:
\begin{equation}\label{relation}
\frac{d_{1}}{s+d_{2}} =\frac{s+d_{1}}{d_{2}}.
\end{equation}
Then using the normalization condition,  Eqs. (\ref{normp1}) and (\ref{normp2}) one can show that
\begin{equation}\label{norm3}
\sum_{n=-\infty}^{\infty} \widetilde{P}_{n,1}(s) = \frac{s+d_{2}}{s(s+d_{1}+d_{2})}
\end{equation}
\begin{equation}\label{norm4}
\sum_{n=-\infty}^{\infty} \widetilde{P}_{n,2}(s) = \frac{d_{1}}{s(s+d_{1}+d_{2})}
\end{equation}
Next we make an assumption that 
\begin{equation}\label{assume}
\widetilde{P}_{n,2}(s) = K \widetilde{P}_{n,1}(s)
\end{equation}
and from Eqs. (\ref{norm3}) and (\ref{norm4}) the unknown parameter $K$ can be easily determined,
\begin{equation}\label{const}
K=\frac{d_{1}}{s+d_{2}}=\frac{s+d_{1}}{d_{2}}.
\end{equation}
In terms of waiting-time distributions this expression can be rewritten as
 \begin{eqnarray}\label{rela}
\frac{\widetilde{\psi}^{\gamma}(s)}{1-\widetilde{\psi_{1}}^{+}(s)-\widetilde{\psi_{1}}^{-}(s)}
=\frac{\widetilde{\psi}^{\delta}(s)}{1-\widetilde{\psi_{2}}^{+}(s)-\widetilde{\psi_{2}}^{-}(s)}.
\end{eqnarray}
This is an important result since it sets a constraint on possible waiting-time distribution functions that describe homogeneous CTRW on the coupled parallel lattices.

Combining Eqs. (\ref{recurp1}) and (\ref{recurp2}) with (\ref{const}), we obtain:
\begin{equation}\label{eqP}
A \widetilde{P}_{n,1}(s)=\delta_{n,0}+ B \widetilde{P}_{n-1,1}(s) + C \widetilde{P}_{n+1,1}(s),
 \end{equation}
where new auxiliary functions are defined as
\begin{eqnarray}
A \equiv s+b_{1}+c_{1}+\frac{d_{1}(s+b_{2}+c_{2})}{s+d_{2}}, \nonumber \\
B \equiv s+b_{1}+\frac{d_{1}b_{2}}{s+d_{2}}, \nonumber \\
C \equiv s+c_{1}+\frac{d_{1}c_{2}}{s+d_{2}},
\end{eqnarray}  
The recursion relation (\ref{eqP}) is similar to Eq. (\ref{eqB}), and the solution again can be found in the following form,\cite{kolomeisky09}
\begin{equation}\label{eqPar}
\widetilde{P}_{n,1}(s) = \left(\frac{B}{C}\right)^{n/2} \left( \frac{ A- \sqrt{A^{2}-4 BC}}{2 \sqrt{BC}}\right)^{|n|} \frac{1}{\sqrt{A^{2}-4BC}}.
\end{equation}
Similar expression for $P_{n,2}(s)$ can be easily obtained from Eqs. (\ref{assume}) and (\ref{const}).

The Eq. (\ref{eqPar}) provides a full description of the CTRW dynamics on the coupled parallel channels at all times. It will be illustrated by looking at first two moments of motion. It can be shown that the Laplace transform of the first moment is equal to
 \begin{equation}\label{eqns}
\langle \widetilde{n}(s) \rangle= \sum_{n=-\infty}^{\infty} n( \widetilde{P}_{n,1}(s)+ \widetilde{P}_{n,2}(s) ) = (1+\frac{d_{1}}{s+d_{2}}) \sum_{n=-\infty}^{\infty} n \widetilde{P}_{n,1}(s),
\end{equation}
which can also be written as 
\begin{equation}\label{eqns2}
\langle \widetilde{n}(s) \rangle=\frac{(b_{1}-c_{1})(s+d_{2})+(b_{2}-c_{2})(s+d_{1})}{s^{2}(s+d_{1}+d_{2})}.
\end{equation}
Utilizing waiting-time distributions it can be shown that,
\begin{eqnarray}\label{eqnns}
\langle \widetilde{n}(s)\rangle=\frac{\left( \widetilde{\varphi_{1}}^{+}(s)-\widetilde{\varphi_{1}}^{-}(s) \right) \left(s+ \widetilde{\varphi}^{\delta}(s)\right)+\left(\widetilde{\varphi_{2}}^{+}(s)-\widetilde{\varphi_{2}}^{-}(s)\right)\left(s+ \widetilde{\varphi}^{\gamma}(s)\right)}{s^2 \left(s+\widetilde{\varphi}^{\gamma}(s)+\widetilde{\varphi}^{\delta}(s)\right)}.
\end{eqnarray}
Similar calculations for the Laplace transform of the second moment produce
\begin{eqnarray}\label{eqnn2s}
 & \langle \widetilde{n^2}(s)\rangle  = \frac{[s+ \widetilde{\varphi}^{\delta}(s)]^2 \left(s[\widetilde{\varphi_{1}}^{+}(s)+\widetilde{\varphi_{1}}^{-}(s)] +2[\widetilde{\varphi_{1}}^{+}(s)-\widetilde{\varphi_{1}}^{-}(s)]^2 \right)}{s^3(s+\widetilde{\varphi}^{\gamma}(s)+\widetilde{\varphi}^{\delta}(s))^2}   \nonumber  &  \\
& + \frac{\widetilde{\varphi}^{\delta}(s)\widetilde{\varphi}^{\gamma}(s)\left( s[\widetilde{\varphi_{1}}^{+}(s)+\widetilde{\varphi_{1}}^{-}(s)+\widetilde{\varphi_{2}}^{+}(s)+\widetilde{\varphi_{2}}^{-}(s)]+4[\widetilde{\varphi_{2}}^{+}(s)-\widetilde{\varphi_{2}}^{-}(s)][\widetilde{\varphi_{1}}^{+}(s)-\widetilde{\varphi_{1}}^{-}(s)] \right)}{s^3(s+\widetilde{\varphi}^{\gamma}(s)+\widetilde{\varphi}^{\delta}(s))^2}  \nonumber  &  \\ 
&  + \frac{[s+ \widetilde{\varphi}^{\gamma}(s)]^2 \left(s[\widetilde{\varphi_{2}}^{+}(s)+\widetilde{\varphi_{2}}^{-}(s)]+2[\widetilde{\varphi_{2}}^{+}(s)-\widetilde{\varphi_{2}}^{-}(s)]^2 \right)} {s^3(s+\widetilde{\varphi}^{\gamma}(s)+\widetilde{\varphi}^{\delta}(s))^2}.   & 
\end{eqnarray}
The stationary-state behavior ($t \rightarrow \infty$) of the first and second moments of the motion can be found by using expansions of waiting-time rate distribution functions at $s \rightarrow 0$,  
\begin{eqnarray}
&\widetilde{\varphi_{1}}^{+}(s) \simeq u_{1} + g_{1}^{+}s + \cdots, & \widetilde{\varphi_{2}}^{-}(s)  \simeq w_{2} + g_{2}^{-}s + \cdots, \nonumber \\
&\widetilde{\varphi_{1}}^{-}(s) \simeq w_{1} + g_{1}^{-}s + \cdots, & \widetilde{\varphi_{2}}^{+}(s) \simeq u_{2} + g_{2}^{+}s + \cdots, \nonumber \\
&\widetilde{\varphi}^{\delta}(s) \simeq \delta + g^{\delta}s + \cdots, & \widetilde{\varphi}^{\gamma}(s) \simeq \gamma + g^{\gamma}s + \cdots,
\end{eqnarray}
where $u_{i}=\widetilde{\varphi_{i}}^{+}(s=0)$, $w_{i}=\widetilde{\varphi_{i}}^{-}(s=0)$ for $i=1,2$; $\delta=\widetilde{\varphi}^{\delta}(s=0)$ and $\gamma=\widetilde{\varphi}^{\gamma}(s=0)$ are effective transition rates.\cite{kolomeisky00} Other coefficients are given by $g_{i}^{\pm}=\frac{d\widetilde{\varphi_{i}}^{\pm}}{ds}|_{s=0}$ for $i=1,2$, $g^{\delta}=\frac{d\widetilde{\varphi}^{\beta}}{ds}|_{s=0}$ and $g^{\gamma}=\frac{d\widetilde{\varphi}^{\gamma}}{ds}|_{s=0}$. The asymptotic calculations then produce the following results for the average position of the random walker,
\begin{eqnarray}
 & \langle n(t) \rangle \simeq [\frac{(u_{1}-w_{1})\delta+(u_{2}-w_{2})\gamma}{\delta+\gamma}]t  & \nonumber \\
& + \frac{1}{\gamma+\delta} \left[(u_{1}-w_{1})(1+g^{\delta})+\delta(g_{1}^{+}-g_{1}^{-})+(u_{2}-w_{2})(1+g^{\gamma})+\gamma(g_{2}^{+}-g_{2}^{-})\right] & \nonumber \\
& - \frac{(1+g^{\delta}+g^{\gamma})[\delta(u_{1}-w_{1})+\gamma(u_{2}-w_{2})]}{(\gamma+\delta)^2}. & 
\end{eqnarray}
The expression for mean-squared position of the particle is more complex,
\begin{eqnarray}
 & \langle n^2(t) \rangle \simeq [\frac{(u_{1}-w_{1})\delta+(u_{2}-w_{2})\gamma}{\delta+\gamma}]^{2}t^{2}   &  \nonumber \\
& + \left \{ \frac{[(u_{1}+w_{1})\delta +(u_{2}+w_{2})\gamma]}{\delta+\gamma}- \frac{2(1+g^{\delta}+g^{\gamma})[\delta(u_{1}-w_{1})+\gamma(u_{2}-w_{2})]^2}{(\gamma+\delta)^3}  \right. & \nonumber \\
& + \frac{\left[ 4 \delta^{2} (u_{1}-w_{1})(g_{1}^{+}-g_{1}^{-})+ 4 \gamma^{2}(u_{2}-w_{2})(g_{2}^{+}-g_{2}^{-})+ 4 \delta \gamma(u_{1}-w_{1})(g_{2}^{+}-g_{2}^{-})+ 4 \delta \gamma(u_{2}-w_{2})(g_{1}^{+}-g_{1}^{-})\right]}{(\delta+\gamma)^2}   & \nonumber \\
& \left. + \frac{\left[ 4 \delta(u_{1}-w_{1})^2(1+g^{\delta})+4 \gamma(u_{2}-w_{2})^2(1+g^{\gamma})+4(u_{2}-w_{2})(u_{1}-w_{1})(\gamma g^{\delta}+\delta g^{\gamma})\right]}{(\delta+\gamma)^2} \right\} t.   &  
\end{eqnarray}
These expressions allow us to derive the stationary drift velocity,
\begin{eqnarray}\label{vp}
  V  \simeq \frac{(u_{1}-w_{1})\delta+(u_{2}-w_{2})\gamma}{\delta+\gamma};
\end{eqnarray}
and the effective diffusion constant, 
\begin{eqnarray}\label{Dp}
& D= \frac{[(u_{1}+w_{1})\delta +(u_{2}+w_{2})\gamma]}{2(\delta+\gamma)} & \nonumber \\
+&  \frac{\left[\delta^{2}(u_{1}-w_{1})(g_{1}^{+}-g_{1}^{-})+ \gamma^{2}(u_{2}-w_{2})(g_{2}^{+}-g_{2}^{-})+ \delta \gamma(u_{1}-w_{1})(g_{2}^{+}-g_{2}^{-})+\delta \gamma(u_{2}-w_{2})(g_{1}^{+}-g_{1}^{-})\right]}{(\delta+\gamma)^2}  & \nonumber \\
& + \frac{\gamma \delta(u_{1}-w_{1})^2(1+g^{\delta})+\delta \gamma(u_{2}-w_{2})^2-\delta^{2} g^{\gamma}(u_{1}-w_{1})^2-\gamma^{2}g^{\delta}(u_{2}-w_{2})^2}{(\delta+\gamma)^3} & \nonumber \\
& - \frac{(u_{1}-w_{1})(u_{2}-w_{2})[\delta \gamma(2+ g^{\gamma}+g^{\delta})+\gamma^{2}g^{\delta}+\delta^{2} g^{\gamma}]}{(\delta+\gamma)^3}.& 
\end{eqnarray}

When the waiting-time distribution functions are exponential all transitions in the system become Poissonian and CTRW models are reduced to simple random walk processes. In this case we have $g_{1}^{+}=g_{1}^{-}=g_{2}^{+}=g_{2}^{-}=g^{\gamma}=g^{\delta}=0$, and the expression for dispersion is much simpler: 
\begin{eqnarray}
D= \frac{[(u_{1}+w_{1})\delta +(u_{2}+w_{2})\gamma]}{2(\delta+\gamma)}+ \frac{[(u_{1}-w_{1})^{2}-(u_{2}-w_{2})^{2}]}{(\delta+\gamma)^3}.
\end{eqnarray}
It agrees exactly with the formula obtained earlier for stationary properties of ordinary random walks on coupled parallel-chain lattices.\cite{kolomeisky06} 

To test the validity of our approach, one could notice that the CTRW model with the branched states can be viewed as a special case of the CTRW on coupled parallel channels if the motion along one of the lattice is disabled, i.e., $\psi_{1}^{\pm}(t)=0$ or $\psi_{2}^{\pm}(t)=0$. Then one can show  that in this case the Laplace transform for the probability function to find the particle at the site $n$, Eq.(\ref{eqPar}), reduces to the main result of Section 2, Eq. (\ref{eq_Pn_phi}), as expected.

\section{Generalized Fluctuation Theorem}

Fluctuation theorems are important for understanding fundamental mechanisms of complex phenomena.\cite{gallavotti95,gaspard04,seifert12} Generalized fluctuation theorems, which reduce to the original formulation under some condition, have been introduced and studied for several CTRW models.\cite{BW08,kolomeisky09} The analysis is performed by considering the ratio of $\frac{\widetilde{P}_{n}(s)}{\widetilde{P}_{-n}(s)}$. Theoretical results obtained in this paper allows us to investigate explicitly generalized fluctuation theorems for CTRW on complex networks.

For CTRW model with branches it can be shown that
\begin{equation}\label{fluctb}
\frac{\widetilde{P}_{n,0}(s)}{\widetilde{P}_{-n,0}(s)}=\frac{\widetilde{P}_{n,1}(s)}{\widetilde{P}_{-n,1}(s)}=\left[ \frac{\widetilde{\psi}^{+}(s)}{\widetilde{\psi}^{-}(s)} \right]^{n}=\left[ \frac{\widetilde{\varphi}^{+}(s)}{\widetilde{\varphi}^{-}(s)} \right]^{n}.
\end{equation}
It is the same formula as for the homogeneous CTRW on lattices without branched states,\cite{BW08,kolomeisky09} suggesting that reversible detachments do not affect the ratio of probabilities  for forward and backward steps of the random walker.  Consequently, branched states  do not change statistics for occurrence of different fluctuations. It leads to the original fluctuation theorem result when $\frac{\psi^{+}(t)}{\psi^{-}(t)}$ is time-independent.

For the CTRW model on coupled parallel channels the results are more complex,
\begin{eqnarray}\label{fluctp}
\frac{\widetilde{P}_{n,1}(s)}{\widetilde{P}_{-n,1}(s)}=\frac{\widetilde{P}_{n,2}(s)}{\widetilde{P}_{-n,2}(s)}=\left[\frac{\widetilde{\varphi_{1}}^{+}(s)(s+\widetilde{\varphi}^{\delta}(s))+\widetilde{\varphi_{2}}^{+}(s)\widetilde{\varphi}^{\gamma}(s)}{\widetilde{\varphi_{1}}^{-}(s)(s+\widetilde{\varphi}^{\delta}(s))+\widetilde{\varphi_{2}}^{-}(s)\widetilde{\varphi}^{\gamma}(s)} \right]^{n}\nonumber \\
=\left[\frac{\widetilde{\varphi_{2}}^{+}(s)(s+\widetilde{\varphi}^{\gamma}(s))+\widetilde{\varphi_{1}}^{+}(s)\widetilde{\varphi}^{\delta}(s)}{\widetilde{\varphi_{2}}^{-}(s)(s+\widetilde{\varphi}^{\gamma}(s))+\widetilde{\varphi_{1}}^{-}(s)\widetilde{\varphi}^{\delta}(s)} \right]^{n}.
\end{eqnarray}
In terms of the waiting-time distribution functions the generalized fluctuations theorem has the following form,
\begin{eqnarray}\label{fluctp1}
\frac{\widetilde{P}_{n,1}(s)}{\widetilde{P}_{-n,1}(s)}=\frac{\widetilde{P}_{n,2}(s)}{\widetilde{P}_{-n,2}(s)}= \left[\frac{\widetilde{\psi_{1}}^{+}(s)(1-\widetilde{\psi_{2}}^{+}(s)-\widetilde{\psi_{2}}^{-}(s))+\widetilde{\psi_{2}}^{+}(s)\widetilde{\psi}^{\gamma}(s)}{\widetilde{\psi_{1}}^{-}(s)(1-\widetilde{\psi_{2}}^{+}(s)-\widetilde{\psi_{2}}^{-}(s))+\widetilde{\psi_{2}}^{-}(s)\widetilde{\psi}^{\gamma}(s)}\right]^{n} \nonumber \\
= \left[\frac{\widetilde{\psi_{2}}^{+}(s)(1-\widetilde{\psi_{1}}^{+}(s)-\widetilde{\psi_{1}}^{-}(s))+\widetilde{\psi_{1}}^{+}(s)\widetilde{\psi}^{\delta}(s)}{\widetilde{\psi_{2}}^{-}(s)(1-\widetilde{\psi_{1}}^{+}(s)-\widetilde{\psi_{1}}^{-}(s))+\widetilde{\psi_{1}}^{-}(s)\widetilde{\psi}^{\delta}(s)}\right]^{n}.
\end{eqnarray}
This result can be understood physically using the following arguments. The numerator can be viewed as an effective waiting-time distribution function to move forward, which is averaged over finding the particle on the lattice 1 and 2. The denominator has the  meaning for effective waiting-time distribution function to move backward. It also shows the importance of transitions between channels on  individual trajectories and on statistics of fluctuations.

\section{Summary and Conclusions}

A theoretical method of calculating all-time dynamics of continuous-time random walks is extended to processes that take place on complex networks. Specifically dynamic properties of CTRW models on the lattices with branched states and CTRW models on the coupled parallel channels are analyzed at all times. The theoretical approach is flexible and robust to deal with complex CTRW models since it is based on the construction of the generalized master equations which are solved exactly in the Laplace space. It suggests that homogeneous CTRW processes on any networks can be analyzed in the similar way.

Our calculations yielded stationary-state dynamic properties for processes that can reach the steady states. All derived results for drift velocities and dispersions of CTRW models on complex networks at stationary conditions agree with available large-time expressions obtained by different methods. In addition, generalized fluctuation theorems are discusses. It is shown that the presence of branched states do not affect fluctuation dynamics of particles, while in the model with coupled parallel lattices transitions between channels are important. It will be interesting to extend this method to more complex inhomogeneous CTRW models on networks\cite{kolomeisky11} that will help to understand better fundamental mechanisms of various complex dynamic phenomena.

\section*{Acknowledgments}

We would like to acknowledge  the support from the Welch Foundation (grant C-1559).

\newpage

\noindent {\bf Figure Captions:} \\
\vspace{5mm}

\noindent Fig. 1. General schemes for continuous-time random walk (CTRW) models on complex networks. a) A lattice model with branched states. Here $\psi_{i}^{+}(t)$, $\psi_{i}^{-}(t)$,  $\psi_{i}^{\beta}(t)$ and $\psi_{i}^{\gamma}(t)$ are waiting-time distribution functions to step forward, backward, to the branched state and out of the branched state, correspondingly.   b) A parallel-chain lattice model.  Here $\psi_{j,i}^{+}(t)$ and $\psi_{j,i}^{-}(t)$ are waiting-time distribution functions to move forward or backward on the lattice $j=1$ or $j=2$. Also, $\psi_{i}^{\delta}(t)$ and  $\psi_{i}^{\gamma}(t)$ are waiting-time distribution functions to transition between lattices.

\newpage

\begin{figure}[ht]
\begin{center}
\unitlength 1in
  
  [a]\resizebox{3.375in}{2in}{\includegraphics{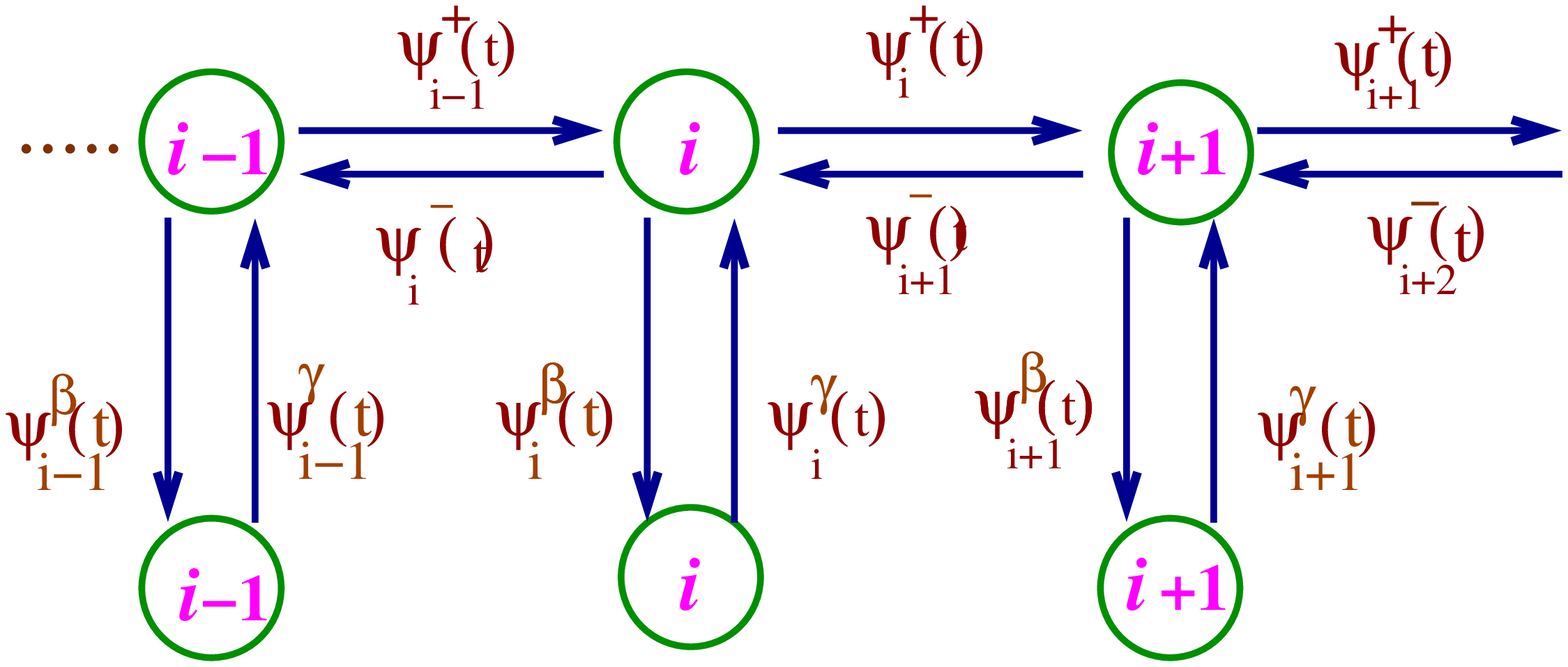}}

  [b]\resizebox{3.375in}{2in}{\includegraphics{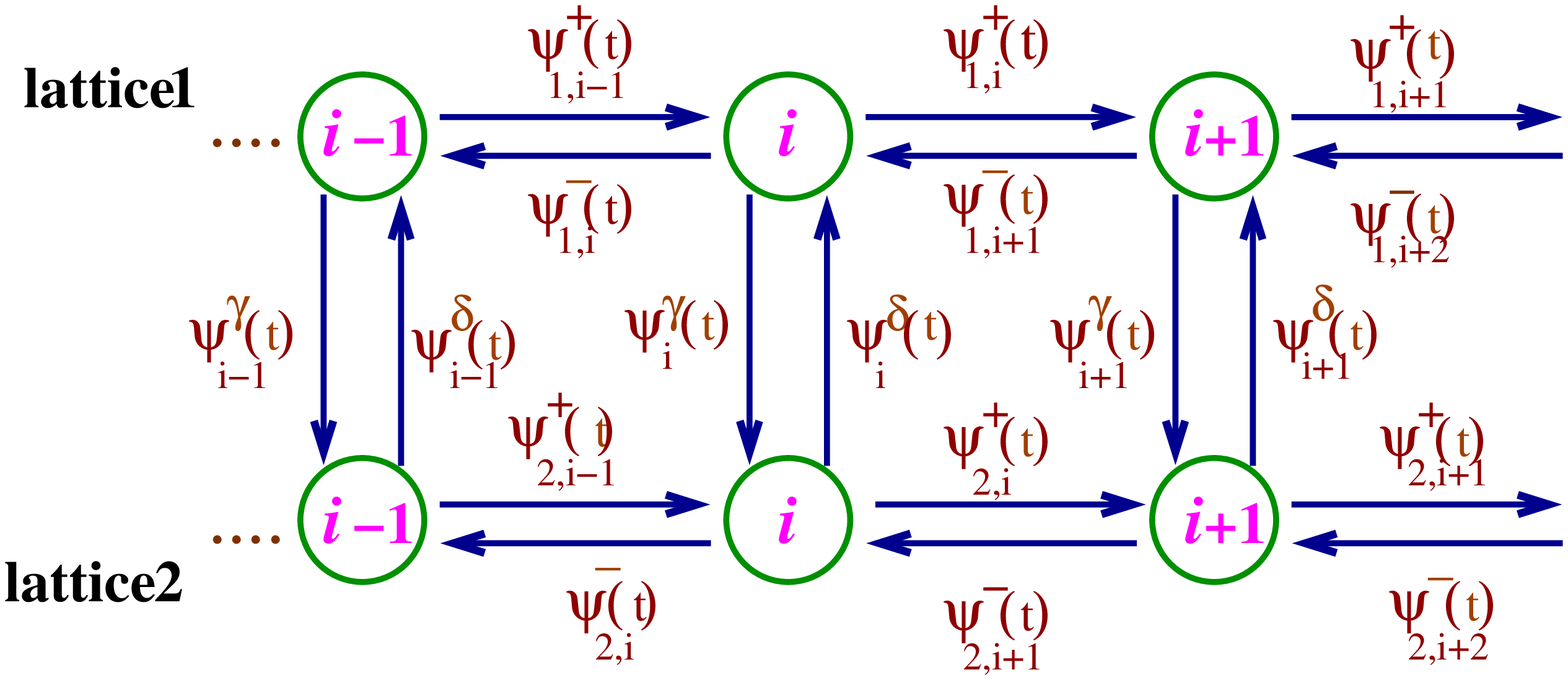}}

\vskip 1in
 \begin{Large} Figure 1. Teimouri and Kolomeisky \end{Large}
\end{center}
\end{figure}

\end{document}